\pdfoutput=1
\documentclass{llncs}

\usepackage{algorithm}
\usepackage{algpseudocode}
\usepackage{caption}
\usepackage{graphicx}
\usepackage{subfigure}
\usepackage[export]{adjustbox}
\usepackage[T1,hyphens]{url}
\begin{document}

\title{Improving Bitcoin's Resilience to Churn}

\author{Nabeel Younis \and Muhammad Anas Imtiaz \and David Starobinski  \and  \\ Ari Trachtenberg}

\institute{
	\email{\{nyounis{,} maimtiaz{,} staro{,} trachen\}@bu.edu} \\
	Boston University{,} ECE Department{,} Boston MA{,} 02215
}

\maketitle
%========== ABS =====================
\begin{abstract} 
Efficient and reliable block propagation on the Bitcoin network is vital for ensuring the scalability of this peer-to-peer network. To this end, several schemes have been proposed over the last few years to speed up the block propagation, most notably the compact block protocol (BIP 152). Despite this, we show experimental evidence that nodes that have recently joined the network may need about ten days until this protocol becomes 90\% effective. This problem is endemic for nodes that do not have persistent network connectivity. We propose to mitigate this ineffectiveness by maintaining mempool synchronization among Bitcoin nodes. For this purpose, we design and implement into Bitcoin a new prioritized data synchronization protocol, called FalafelSync. Our experiments show that FalafelSync helps intermittently connected nodes to maintain better consistency with more stable nodes, thereby showing promise for improving block propagation in the broader network.  In the process, we have also developed an effective logging mechanism for bitcoin nodes which we release for public use. 

% In this paper, we conduct experiments to assess the effectiveness of this protocol. 
% %With the compact block protocol, significant improvements in block propagation times for nodes that are continuously online for several week can be achieved. However, 
% Our experiments show that nodes that have recently joined the network need about a week before the compact block becomes effective [is it all?]. Thus, there is room for improvement for nodes that are missing relevant transactions for the next few blocks. 
% As a potential solution, we propose and implement a prioritized sync protocol that assesses the current mempool and sends out \texttt{txmempoolsync} messages to peers at regular time intervals to sync up the most likely transactions to be included within the next few blocks. Our results show a reduction in the failure rate of compact blocks [need to quantify].
%in a fluctuating node.
%Implementing some form of prioritized transaction synchronization for Bitcoin is critical to help scale Bitcoin and for the success of future projects that depend on mempools being relatively well synced. 
\end{abstract} 

%% Note: Classification and Keywords are only required for the camera-ready version

%
% The code below should be generated by the tool at
% http://dl.acm.org/ccs.cfm
% Please copy and paste the code instead of the example below. 
%
% \begin{CCSXML}
% <ccs2012>
%  <concept>
%   <concept_id>10010520.10010553.10010562</concept_id>
%   <concept_desc>Computer systems organization~Embedded systems</concept_desc>
%   <concept_significance>500</concept_significance>
%  </concept>
%  <concept>
%   <concept_id>10010520.10010575.10010755</concept_id>
%   <concept_desc>Computer systems organization~Redundancy</concept_desc>
%   <concept_significance>300</concept_significance>
%  </concept>
%  <concept>
%   <concept_id>10010520.10010553.10010554</concept_id>
%   <concept_desc>Computer systems organization~Robotics</concept_desc>
%   <concept_significance>100</concept_significance>
%  </concept>
%  <concept>
%   <concept_id>10003033.10003083.10003095</concept_id>
%   <concept_desc>Networks~Network reliability</concept_desc>
%   <concept_significance>100</concept_significance>
%  </concept>
% </ccs2012>  
% \end{CCSXML}
% 
% \ccsdesc[500]{Computer systems organization~Embedded systems}
% \ccsdesc[300]{Computer systems organization~Redundancy}
% \ccsdesc{Computer systems organization~Robotics}
% \ccsdesc[100]{Networks~Network reliability}

% \keywords{ACM proceedings, \LaTeX, text tagging}

%\maketitle

%\sloppy

%============= INTRODUCTION ============
\section{Introduction} 
The Bitcoin cryptocurrency, originally introduced by Satoshi Nakamoto in 2008 \cite{article}, is today an extremely popular peer-to-peer electronic payment system used for buying and selling goods in different markets across the globe. Far beyond its initial purview, today's Bitcoin network has roughly eleven thousand nodes\footnote{In this paper, the term \emph{node} refers to what Bitcoin calls a \emph{full node}, meaning that it implements the full Bitcoin protocol and stores an entire copy of the blockchain.} online at any given time, and this represents a doubling over last year's size~\cite{bitnodes}.  Indeed, together with hundreds of derivative cryptocurrencies, the total market capitalization of these electronic payment systems is roughly half a trillion dollars~\cite{cryptocurrencies}.

Bitcoin's public ledger system is known as \emph{blockchain}, and it records all transactions that take place in the Bitcoin network~\cite{decker2013information}. Each new transaction is broadcast over the network, and thereafter recorded by every node in its local memory pool (called \emph{mempool}) for subsequent consensus-based validation. By design, a new block containing transactions is created (by a \emph{mining} node) and propagated over the network (by the full nodes) roughly once every ten minutes~\cite{bitcoinwikiblock}.  
%When the inter-block arrival rate deviates from ten-minute threshold (as it does in practice), the network agrees to adjust the block creation difficulty toward the designed rate.  
%After a block is approved by other nodes in the Bitcoin network (through mining), it becomes part of the blockchain after a few confirmations.
After a block is successfully mined and accepted by the Bitcoin network, to be cemented in the blockchain, others must mine on top of it.

A key challenge in this context lies in reducing the propagation time of blocks. Indeed, the consequences of slower block propagation times include an increase in the number of ``soft forks'' of the blockchain and a consequent wasting of computational resources~\cite{energy} as Bitcoin miners mine the wrong blockchain head~\cite{decker2013information}.
To address this challenge, the compact block protocol~\cite{BIP152} has been proposed and is currently implemented on the standard Bitcoin Core reference implementation. Another solution, called Graphene~\cite{ozisik2017graphene}, has recently been proposed in the literature. Both of these protocols aim to decrease propagation time to the broader network by reducing the amount of data that needs to be propagated between nodes (see Section~$2$ for more details about these two protocols).

However, like any peer-to-peer network, it is also important that the Bitcoin network be able to support a high rate of \emph{churn}~\cite{stutzbach2006understanding}, that being the rate at which nodes enter and leave the network. Indeed, as the Bitcoin network grows, we can expect the same from the heterogeneity of its constituent nodes. For example, nodes may enter the network from financially restricted or developing countries, with intermittent electricity supply or Internet connectivity; small business and hobbyists may choose to run nodes during off-peak hours, or on cheap, legacy machines; finally, some users may wish to run nodes on mobile devices such as laptops. Therefore, the network must be able to quickly propagate blocks to all current nodes, even as some of these nodes enter or leave the network.

%\footnote{While the average rate of inter-block arrival is designed to be $10$ minutes, experiments show that this rate deviates. For example, blocks created in the year $2017$ had an average rate of $9.39$ minutes. When blocks are consistently mined in under $10$ minutes, the network agrees to adjust the difficulty of the solution in order to bring it back to being roughly $10$ minutes to mine a block.}, a new block containing transactions is created and propagated over the network by the node that creates it. After verification from other nodes in the network, the new block becomes part of the blockchain.
% Anas - I am adding an image below. The mean of the data points is 9.2835 minutes. Data is over the past 5 years period. Kindly add to document if feasible
%\begin{figure}[t]
%	\centering
%    \includegraphics[width=0.48\textwidth]{averageblockminetime.jpg}
%    \caption{Average time to mine a block in minutes (based on data from~\cite{avgblockminetime})}
%    \label{fig:averageblockminetime}
%\end{figure}

The aim of this work is to experimentally evaluate the performance of Bitcoin in the presence of node churn, identify practical weaknesses, and propose an improvement that promises to help block propagation protocols reach their full potential. We first conduct an experiment on the recovery (convergence) time of a Bitcoin node joining the network. The experiment shows it may require as much as ten days to stably attain a 90\% effectiveness of the compact block protocol, even with full uptime and connectivity. Next, in a second set of experiments, we show that when a node is accessible $90\%$ up-time of the time (i.e., repeatedly, $9$ minutes connected, $1$ minute down), the effectiveness of the compact block protocol drops by roughly 50\%, meaning that the node will frequently request additional information from its peers after receiving a new compact block, and this may cause additional delays in block propagation. 

To mitigate this problem, we propose a peer-to-peer prioritized synchronization protocol, dubbed \emph{FalafelSync}, that periodically synchronizes transactions that are most likely to be included in upcoming blocks. We implement FalafelSync within Bitcoin core and repeat the second set of experiments. Our experiments show that FalafelSync dramatically reduces the number of missing transactions in the memory pool of an intermittently connected node, thus leading to superior performance by the Compact block and Graphene protocols.

One infrastructural foundation of our experiments is a novel logging system, which we have released for public use~\cite{ourbitcoinrepo} and should be of independent interest to researchers conducting measurements on Bitcoin. %\footnote{\texttt{/src/logFile.*}} .
This system enables the recording of a variety of propagation-related data, including the state of the mempool, block metadata, and transactions. We have extensively used this system to understand Bitcoin Core, debug FalafelSync and record useful data for different experiments.

\subsection{Contributions}
Our main contributions are thus:
\begin{itemize}
\item Identifying and demonstrating the real impact of churn on the Bitcoin network.
\item Proposing, designing, and developing a prioritized synchronization protocol, called \emph{FalafelSync}, that integrates seamlessly into existing protocols.
\item Implementing our approach on an active Bitcoin Core node, and analyzing the results.
\item Developing a publicly available logging mechanism for analyzing Bitcoin node functionality.
\end{itemize}

% In our experiments, our approach produces a reduction in the failure rate of compact blocks and Graphene blocks, and this
% is particularly noticeable for intermittently connected nodes.

\subsection{Roadmap}
The rest of this paper is organized as follows: In Section~\ref{sec:related}, we cover background and related work. Next, Section~\ref{sec:implementation} describes implementation of a proof-of-concept prioritized transaction synchronization protocol, together with our new log file system for Bitcoin Core. Section~\ref{sec:experiments} provides details on experiments and results, and discussion of their limitations. Finally, we provide a conclusion and directions for future work in Section~\ref{sec:conclusions}.

% =================== RELATED WORK ==========================
\section{Background and Related Work}
\label{sec:related}
We next explain some background technical material relevant to the Bitcoin network, followed by work that is related to our results.

\subsection{Blocks and the Mempool}

\begin{figure}[t]
	\centering
    \makebox[\textwidth][c]{\includegraphics[width=1.2\textwidth]{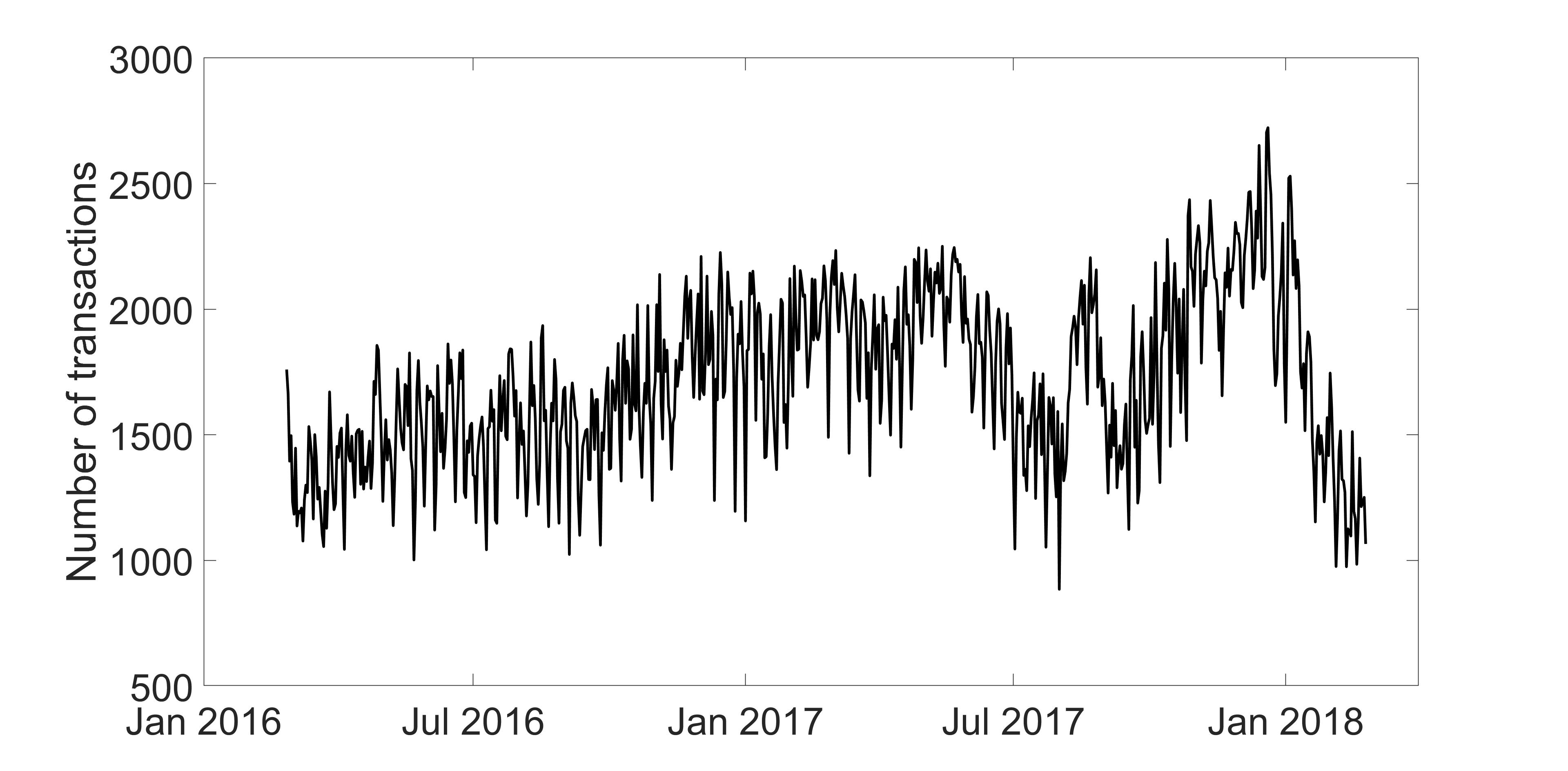}}
    \caption{Average number of transactions per block (based on data from~\cite{avgtxs})}
    \label{fig:numtxs}
\end{figure}

Bitcoin's primary record-keeping mechanism is the \emph{block}. It is a data structure that contains metadata about the block's position in the blockchain together with a number of associated transactions (typically a couple thousand - see Figure~\ref{fig:numtxs})~\cite{avgtxs}. A block is generated roughly every ten minutes through the \emph{mining} process, and, once generated, it and its transactions become a part of the Bitcoin blockchain. There is a probability that different nodes will incorporate different blocks in their version of the blockchain (a process known as a \emph{soft fork}), but these differences are reconciled over time in a competitive process.

In the interim time between when a transaction is announced and when it is included in a block, transactions are stored locally in the \emph{mempool}. The mempool is a constantly changing dataset that stores all the unconfirmed transactions waiting to be included in future blocks. It can contain anywhere between $10{,}000$ and $100{,}000$ transactions, depending on network activity. Currently the mempool experiences between $3$ and $10$ insertions per second~\cite{txspersec}; the arrival of a new block also instigates many deletions from the mempool, between $1000$ and $3000$ transactions every $10$ minutes on average.

\subsection{Block Propagation on the Bitcoin network}

Block Propagation is the process of communicating a newly mined block to the network. It is the backbone of Bitcoin's ability to maintain consensus on the current balances of address (wallets). When a new block is discovered, each Bitcoin node actively immediately advertises the block to all of its neighboring peers. 

\begin{figure}[t]
	\centering
    \makebox[\textwidth][c]{
    \begin{minipage}{.8\textwidth}
   	\centering 
  \includegraphics[width=0.8\linewidth]{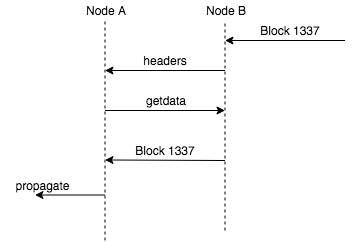}
    %\caption{Legacy Block Propagation~\cite{protodocs}}
    \captionof{figure}{Legacy Block Propagation~\cite{protodocs}}
    \label{fig:normalProp}
   \end{minipage}%
	\begin{minipage}{.8\textwidth}
	\centering
    \includegraphics[width=0.8\linewidth]{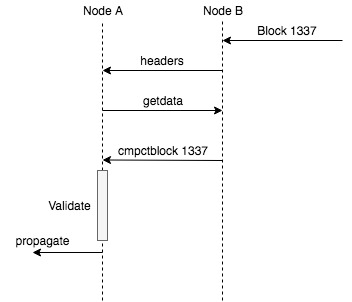}
    %\caption{Successful Compact Block~\cite{BIP152}} 
    \captionof{figure}{Successful Compact Block~\cite{BIP152}}
    \label{fig:cmpctSucc}
    \end{minipage}%
    }
\end{figure}

There are currently two main block protocols in Bitcoin: the original protocol developed for the first implementation of Bitcoin (Figure~\ref{fig:normalProp} and the Compact Block Protocol - BIP 152- (Figure~\ref{fig:cmpctSucc})~\cite{BIP152}. The original protocol is adequate for block propagation but it may require significant network resources - typically close to $1$ MB per block~\cite{blocksize}, making it susceptible to network bottlenecks, especially in less developed areas of the world. 

The \emph{compact block} was developed in an effort to reduce the total bandwidth required for block propagation. As the name implies, the compact block is able to communicate all the necessary data for a node to reconstruct and validate one standard block. The compact block contains the same metadata as the normal block, but instead of sending a full copy of each transaction included, it sends only hashes of transactions. It is important to highlight that the compact block's main feature is sending hashes of transactions in place of a complete copies of transactions. A transaction is between $500$ and $800$ bytes, whereas the hashes used for the compact block are only 6 bytes per transaction, a significant bandwidth saving that relies on the assumption that the receiving node already has the relevant transactions and just needs to know in which blocks they belong. This trade-off makes the compact block magnitudes smaller in size than the original block at the cost of potentially needing extra round-trip communications for transactions whose hashes the receiving node does not recognize.

\begin{figure}[h]
	\centering
    \includegraphics[width=0.65\textwidth]{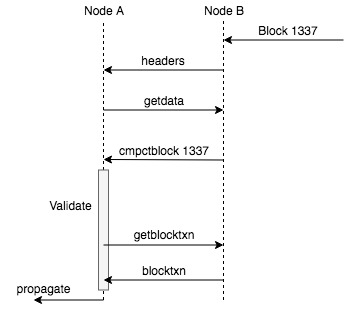}
    \caption{Unsuccessful Compact Block~\cite{BIP152}} 
    \label{fig:cmpctFail}
\end{figure}

If a receiving node's mempool contains all the transactions whose hashes are contained in a compact block that it received, then it will be able to successfully reconstruct the original block. However, if not all transactions are already in the node's mempool then it will fail to reconstruct the block. When the compact block protocol fails, the extra round trips required slow down block propagation from the node it failed at, this can be seen in Figure~\ref{fig:cmpctFail}. If a significant number of nodes are failing to receive the compact block then this could lead to an increase in soft forks and add to the economic and environmental damage caused by miners working on old blocks.

\subsection{Graphene Protocol}

As the network grows, communication complexity plans an increasingly significant role. In an effort to minimize communication, Ozisik \textit{et al.} have developed the Graphene protocol~\cite{ozisik2017graphene}, a new block protocol that couples an Invertible Bloom Lookup Table (IBLT)~\cite{goodrich2011invertible} with a Bloom filter in order to send transaction hashes in a smaller package. Rather than sending each transaction's hash individually, as is done in compact blocks, the Graphene protocol inserts all the transactions hashes into a Bloom filter and an IBLT which are then sent to peers. When a Graphene block is received, it is subtracted from the mempool's IBLT and the rest of the transactions hashes are decoded from the remaining IBLT.  According to the authors in~\cite{ozisik2017graphene}, a Graphene block is capable of being around a fifth of the size of a compact block, and
they provide simulations (but not an actual implementation) demonstrating their system.

\begin{figure}[H]
	\centering
    \includegraphics[width=0.65\textwidth]{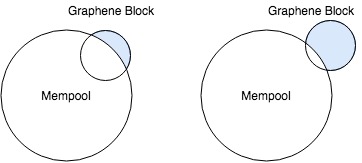}
    \caption{Graphene Block: Left: Successful decoding of IBLT; Right: Unsuccessful decoding of IBLT} 
    \label{fig:grapheneVenn}
\end{figure}

IBLT's provided probabilistic guarantees of decodability, and the authors briefly discuss a fallback procedure in the case of decoding failure. The proposed solution for a failed Graphene block is to send another Graphene block with twice the number of cells in the IBLT. This is a simple solution that addresses decoding the IBLT with a greater probability of success but not one that considers the effect this has on block propagation. Consider the scenario where a node does not have a significant percentage of transactions of a new block (i.e: $15\%$ or more depending on the parameters used) in its mempool, visualized on the right side of Figure~\ref{fig:grapheneVenn}. This node will receive the Graphene block and attempt to decode it. If it fails to do so, it will keep requesting a larger IBLT until it decodes the block successfully. Then, it requests for the missing transactions and only then is finally able to reconstruct the new block, verify it and propagate it to its peers. The previous scenario would require at least $3$ sets of extra messages to successfully reconstruct the block before continuing propagation. This is already a whole round trip slower than the compact block protocol's recovery scheme but due to the IBLT being small compared to just a list of transaction hashes there is a bandwidth saving compared to the compact block's total bandwidth usage.

Another concern with the Graphene simulation is that it does not address the typical case where a given transaction is not in the mempool of a
receiving node.  Indeed, the simulations presented in the Graphene paper assume that:
\begin{itemize}
\item All nodes have very high interconnectivity with $100\%$ uptime.
\item All nodes have updated and highly synchronized mempools.
\item Block propagation time is $2.5$ minutes (as compared to the roughly $10$ minute average for Bitcoin).
\end{itemize}
As a result, it is not clear that the Graphene simulations accurately reflect the current Bitcoin network transients, including the percentage of nodes that are intermittently connected or have unsynchronized mempools.

\begin{table}[b]
\centering
\caption{Comparison of Block Protocols}
\label{table:blocks}
\begin{tabular}{|c|c|c|c|}
\hline
                  & \textbf{Satoshi}   & \textbf{Compact}              & \textbf{Graphene}                             \\ \hline
\textbf{Size in KB}        & $1000$      & $21$                   & $2.6-5.2$                              \\ \hline
\textbf{TX Form}           & Full Copy & List of Hashes       & IBLT of Hashes                       \\ \hline
          &        &                         & Request larger \\
\textbf{Failure}   & N/A    & \texttt{getblocktxn/}   & IBLT than \\ 
\textbf{Procedure} &        & \texttt{blocktxn}       & \texttt{getblocktxn/} \\               &        &                         & \texttt{blocktxn} \\\hline
\end{tabular}
\end{table}

Table~\ref{table:blocks} summarizes the various existing block protocols and their properties.  In the coming sections, we will elaborate on the limitations of these protocols, most notably that their effectiveness relies on the synchronization of mempools among participating nodes. Our approach involves building a synchronizing middleware into the Bitcoin protocol in order to satisfy this assumption and allow both protocols to more fully exhibit their intended benefits.

%================ IMPLEMENTATION ======================

\begin{figure*}[tb]
	\centering
    \subfigure[Log file entry for a successful compact block] {
      \includegraphics[width=1\textwidth]{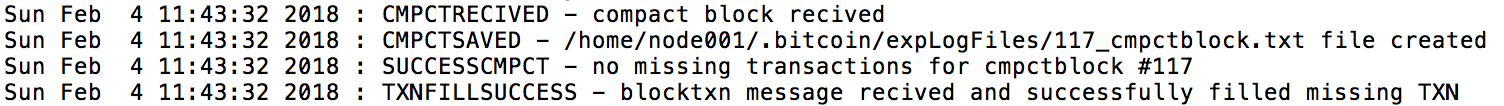}
      \label{fig:successLog}
    }
	\subfigure[Log file entry for a failed compact block] {
      \includegraphics[width=1\textwidth]{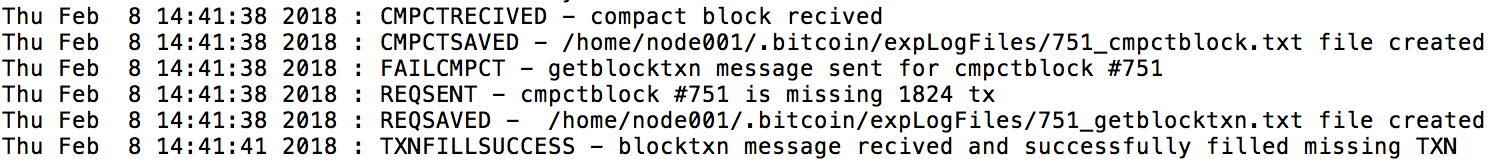}
      \label{fig:failLog}
	}
	\subfigure[Log file entry containing indexes for missing transactions that were requested on receiving a failed compact block] {
      \includegraphics[width=1\textwidth]{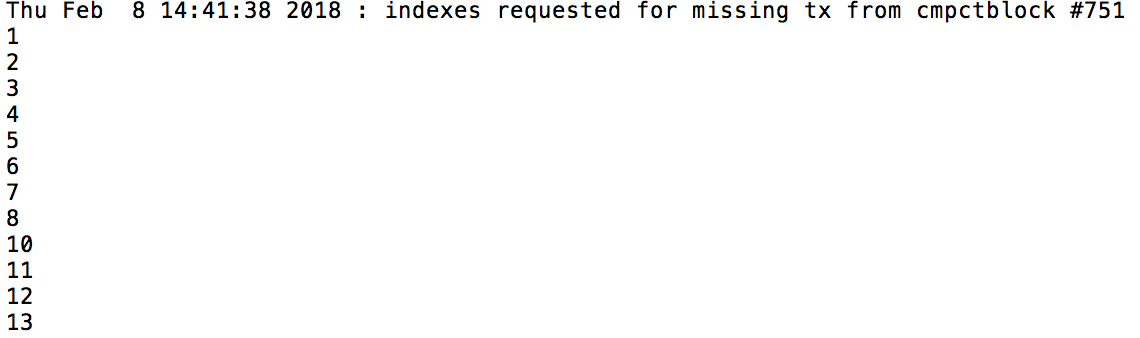}
      \label{fig:failcmptLog}
    }
    \caption{Examples of entries in log files for different events}
    \label{fig:logfileexamples}
\end{figure*}

\section{Implementation and logging}
\label{sec:implementation}
In this section, we introduce the implementation of our logging system, which has been used for observing the behavior of Bitcoin core, and our prioritized transaction synchronization protocol, FalafelSync.

\subsection{Data Collection Mechanism} %previously: Use of Logging System, ALT: logFile, our human-friendly logs
%Our first challenge was how to monitor the behavior of a Bitcoin node and record data from it in real-time as messages came in at the rate of dozens per second. 
A Bitcoin node can experience dozens of messages per second, and meaningfully monitoring the behavior of a Bitcoin node in real-time can thus be a challenge. Worse yet, the Bitcoin Core client (v0.15.0)~\cite{bitcoincore} contains highly interdependent code that makes tracing the code tedious. In effect, it behaves as one large state-machine with several threads calling functions from many different files simultaneously. Fortunately, there does exist a JSON-RPC~\cite{bitcoindevref} for Bitcoin Core's client that is able to fetch us information about the current state of the mempool and its transactions and we have been able to use its \texttt{getrawmempool} and \texttt{getmempoolinfo} calls for collecting preliminary data. However, gathering data about messages of interest such as \texttt{cmpctblock}, \texttt{getblocktxn}, and \texttt{blocktxn} \cite{protodocs}, requires a logging system with a finer granularity for recording data than what the JSON-RPC calls offers.

%But to really understand how the node was behaving, we needed to record its behavior as important messages like \texttt{cmpctblock}, \texttt{getblocktxn}, and \texttt{blocktxn} \cite{protodocs} came in.

To aid in understanding Bitcoin Core's behavior and in debugging our implementation of prioritized transaction synchronization, we have developed a new log-to-file system (\texttt{/src/logFile.*})~\cite{ourbitcoinrepo} that produces human-friendly, easy-to-read text files. This new logging mechanism allows us to isolate specific behaviors through select calls anywhere within the Bitcoin core's source code, most notably information about different protocols such as the compact block and our sync protocol. Our logging system writes core data to a log file, and also can record various events and the information associated with those events. For instance, when a compact block arrives, we log it and save the transaction hashes included in the compact block in a separate file with a unique identifier tying it to a log entry (as seen in Figure~\ref{fig:logfileexamples}). We have used this system as our primary data collection mechanism for all of our experiments. 

\subsection{FalafelSync}

\begin{figure}[tb]
	\centering
    \includegraphics[width=0.5\textwidth]{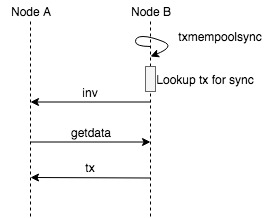}
    \caption{FalafelSync~\cite{ourbitcoinrepo}} 
    \label{fig:falafelsync}
\end{figure}

Bitcoin Core's source code's complexity and its constant updating means there is a minimal amount of current documentation on the code itself, and nearly no documentation on how to add new functionality to Bitcoin Core. Among the resources we found were outdated Bitcoin Wiki pages, conversations between contributors on GitHub pull requests, bug reports, and the Bitcoin Improvement Proposals (BIPs). This environment conspired to increase the complexity of implementing FalafelSync. Indeed, implementation of FalafelSync involved a few months of reverse engineering the Bitcoin Core networking source code with the help of our new logging system.

Within Bitcoin's source code we have added the \texttt{txmempoolsync} protocol that packages within one inventory (\texttt{inv}) message the larger number of transactions between the top $10\%$ of transactions and the top $1000$ transactions in the mempool sorted according to their \texttt{ancestor-score}, an internal scoring mechanism of Bitcoin that ranks transactions within according to the total transaction fees of the transactions and all of its ancestors that are still unconfirmed.  In essence, the \texttt{inv} message provides our sync-related metadata. The rest of the sync process then involves sifting through what is already in the receiving node's mempool and requesting the missing transactions is already handled automatically with Bitcoin Core's \texttt{getdata} and \texttt{tx} messages.  

\begin{algorithm}[t]
	\caption{Steps in FalafelSync}
    \begin{algorithmic}[1]
    	\State Send self \texttt{txmempoolsync} message
        \State Access the top transactions ranked by ancestor score
        \State Package transaction hashes into an \texttt{inv} message
        \State Send \texttt{inv} message to all peers
        \State Return control to Bitcoin Core
    \end{algorithmic}
\end{algorithm}

In the following experiments, we use two nodes to test FalafelSync. To ensure that both nodes were connected to each other, we utilized a python script that sends an \texttt{addnode}\footnote{\texttt{addnode} is a Bitcoin Core JSON-RPC call that creates a connection to the specified IP address} call to Bitcoin Core. 
%This way, we ensured that if both nodes were online and the sync message is sent, then it would reach the node we want it to. 

\subsubsection{Pseudo-periodic triggering of FalafelSync}
Within Bitcoin Core there is no internal timing mechanism for triggering a sync protocol at precise, regular, time intervals. As such, our triggering mechanism adds a pseudo-timer that counts the number of incoming messages. When it reached the message count trigger limit\footnote{To set the trigger limit we took the average number of incoming messages per second and multiplied to get a trigger limit that takes roughly ten minutes to reach.}, we would interrupt the next incoming message and internally send a \texttt{txmempoolsync} message to the network message processing function that would then execute the FalafelSync protocol. After the FalafelSync is completed Bitcoin Core returns to process the interrupted message from earlier. 

With this basic implementation of a sync protocol within Bitcoin Core, we have laid the foundations for future work on implementing more sophisticated synchronization protocols based on the existing sync literature~\cite{minsky2003set,goodrich2011invertible}. %maybe move to conclusion 

\begin{figure*}[t]
	\centering
    \includegraphics[width=0.9\textwidth]{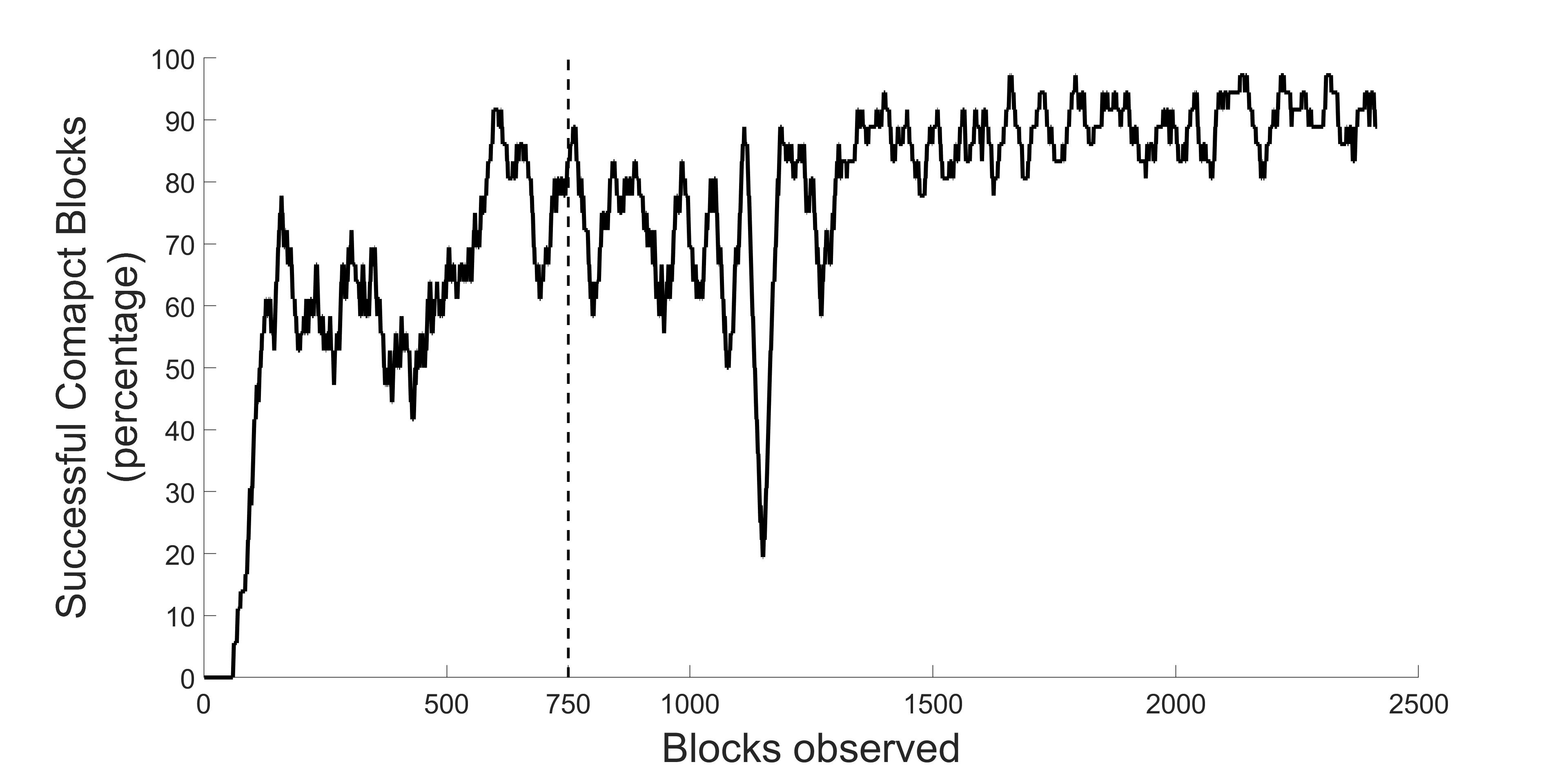}
    \caption{Compact block success rate during high network activity}
    \label{fig:old_experiment_data}
\end{figure*}
%===================== EXPERIMENTAL RESULTS ================
\section{Experimental Results}
\label{sec:experiments}

% The following experiments  nodes set up in newer parts of the network and demonstrate the limitations of the compact block protocol with these nodes.

% \subsection{Experiments Road Map}

% In our first experiment, we aimed to document how long it would take a node to recovery from a recent restart with an "empty" mempool. Following our implementation of FalafelSync, we conducted two experiments: the first shows the current behavior of a node with an intermittent network connection which was recently restarted. The second one tests how well FalafelSync works. 

In this section, we report experiments on the performance of the compact block protocol in Bitcoin. In our first experiment, we measure, over a long period time, the recovery of a node joining the Bitcoin network with an empty mempool. The second and third experiments consider two nodes joining the Bitcoin network. One node has stable connectivity and the other has intermittent connectivity (i.e., it is periodically ``on'' for 9 minutes, then ``off'' for 1 minute, and so on). 

We first run an experiment without FalafelSync (second experiment) and then one with FalafelSync (third experiment). Note that the FalafelSync protocol runs between the stable node and the intermittent node, when the latter one is ``on''. Finally, we report simulations using the experimental data to evaluate the potential performance improvement that FalafelSync could yield for the Graphene protocol. Our experiments were run on during the following date ranges:
\begin{itemize}
\item First Experiment: Aug 23, 2017 - Sep 8, 2017 
\item Second Experiment: Feb 3, 2018 - Feb 9, 2018
\item Third Experiment: Feb 16, 2018 - Feb 22, 2018
\end{itemize}

\subsection{Recovery Time of a Node Joining the Network}
In this first experiment, we measure the performance of the compact block protocol when a new node joins the network with an ``empty'' mempool. An empty mempool is not necessarily a mempool with no transactions but also one that contains outdated transactions loaded from an old \texttt{/.bitcoin/mempool.dat} file\footnote{The \texttt{mempool.dat} file is created when a node is shutting down, it stores the entire mempool at the time of the shutdown. Upon restarting a node the mempool.dat file is reloaded into the mempool then deleted.}. 

% We wanted to observe how long it would take a recently restarted node to recover from its downtime. For the purposes of our analysis of the results, recovery time is the amount of time it takes a node to receive compact blocks with a stable $90$ percent success rate (i.e receiving a compact block without requesting missing transactions) in a given window. 

This experiment is performed on an Ubuntu 16.04 LTS machine running Bitcoin Core v0.15.0. 
% The measurement data is collected using the logging system described earlier. 
 Prior to the start of the experiment, the node is on the network for $24$ hours before being turned off for $24$ hours to create an ``empty'' mempool with outdated transactions. The logging system described earlier is used to collect data on the status of compact blocks and the number of missing transactions.

Figure~\ref{fig:old_experiment_data} shows the first two weeks of performance after a node joins the system. The x-axis in the graph displayed in Figures~\ref{fig:old_experiment_data}, \ref{fig:compsuccessrate_withoutsync}, \ref{fig:compsuccessrate_withsync}, and \ref{fig:comp_node003} use block-time units (number of blocks observed since the start of the experiment). 
% Note that the blockchain has no concept of human-time, only what block came before so all measurements were taken over block-time. 
For reference, $144$ blocks roughly translate into $24$ hours. The y-axis represents the success rate of the compact block protocol, which is a moving average computed over the past 36 blocks. Note that the vertical line in Fig.~\ref{fig:old_experiment_data} is used to correlate this figure with ones to follow which have results for $750$ blocks each.

The figure shows that it takes for the compact block protocol about $400$ (\textasciitilde$3$ days) blocks to reach a 50\% success rate and about $1500$ blocks (\textasciitilde$10$ days) to reach a 90\% success rate. However, even after a $50\%$ success rate is reached, a node can still miss many blocks (e.g., between block-time $1100$ and $1200$). 

Over the course of this experiment, we do observe brief time periods during which the node is accepting compact blocks at a $100\%$ success rate, but this level of performance is not sustained. This could be due to a combination of transactions not reaching the node and Bitcoin's directed connectivity. Specifically, Bitcoin's network is made up of directed connections, hence it is not guaranteed that every incoming connection is paired with an outgoing connection. The node used in this experiment usually had between 8 and 12 outgoing connections and 10 to 20 incoming connections.

The combination of slow recovery time and randomly missing transactions significantly affects the effectiveness of the compact block protocol. These results also indicate that Bitcoin is unable to effectively support a high churn rate.

It should be noted that this experiment was conducted in Fall 2017, a time at which the Bitcoin network was going under a significant amount of stress as it struggled to handle the high volume of transactions.

\subsection{Intermittent Network Connectivity without FalafelSync}
\label{subsec:intermittent}

\begin{figure}[!b]
	\centering
    \makebox[\textwidth][c]{\includegraphics[width=1.2\textwidth]{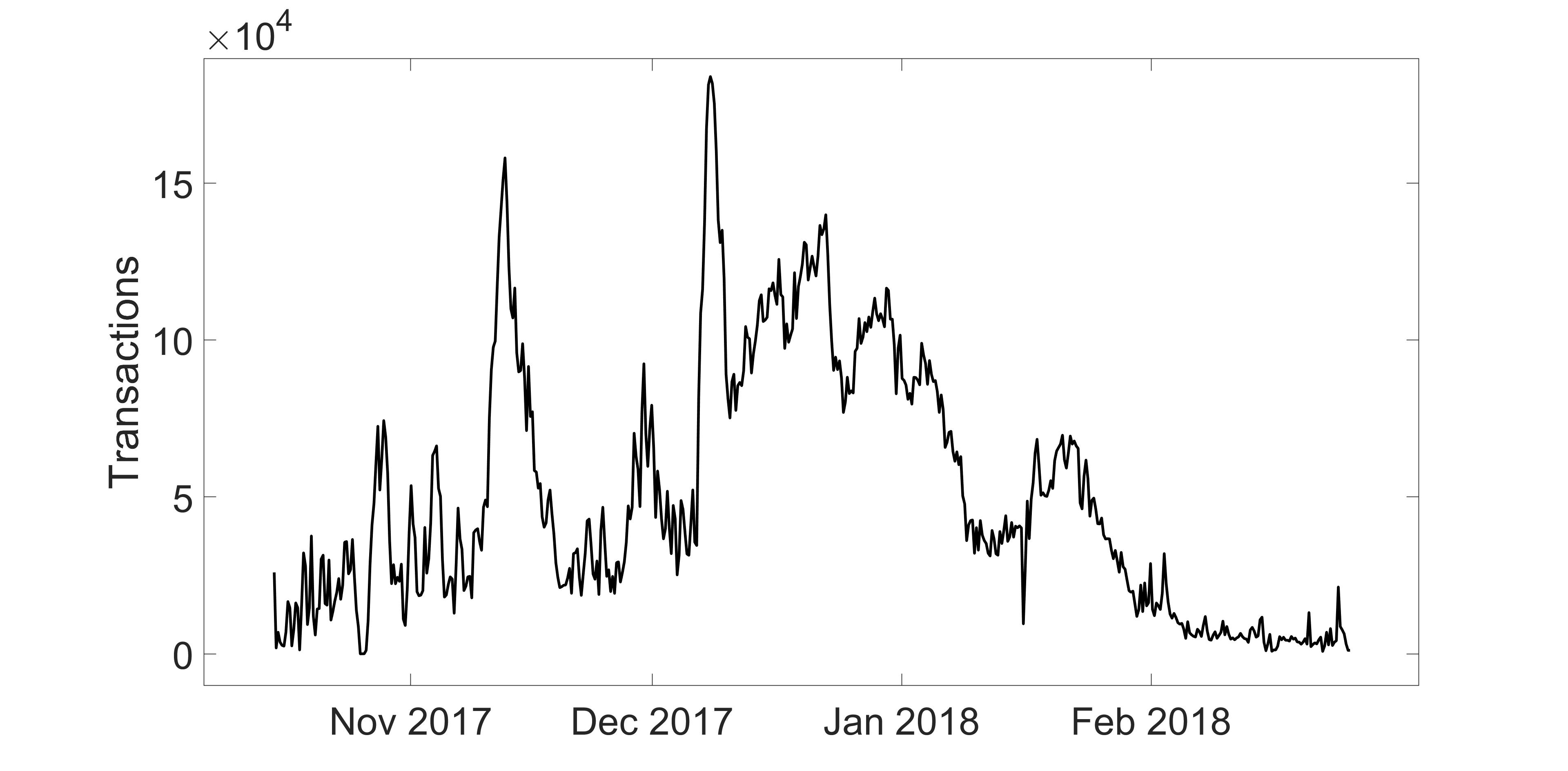}}
    \caption{Mempool transaction count (based on data from \cite{mempooltxs})}
    \label{fig:mempooltxs}
\end{figure}

\begin{figure}[t]
	\centering
    \makebox[\textwidth][c]{\includegraphics[width=1.2\textwidth]{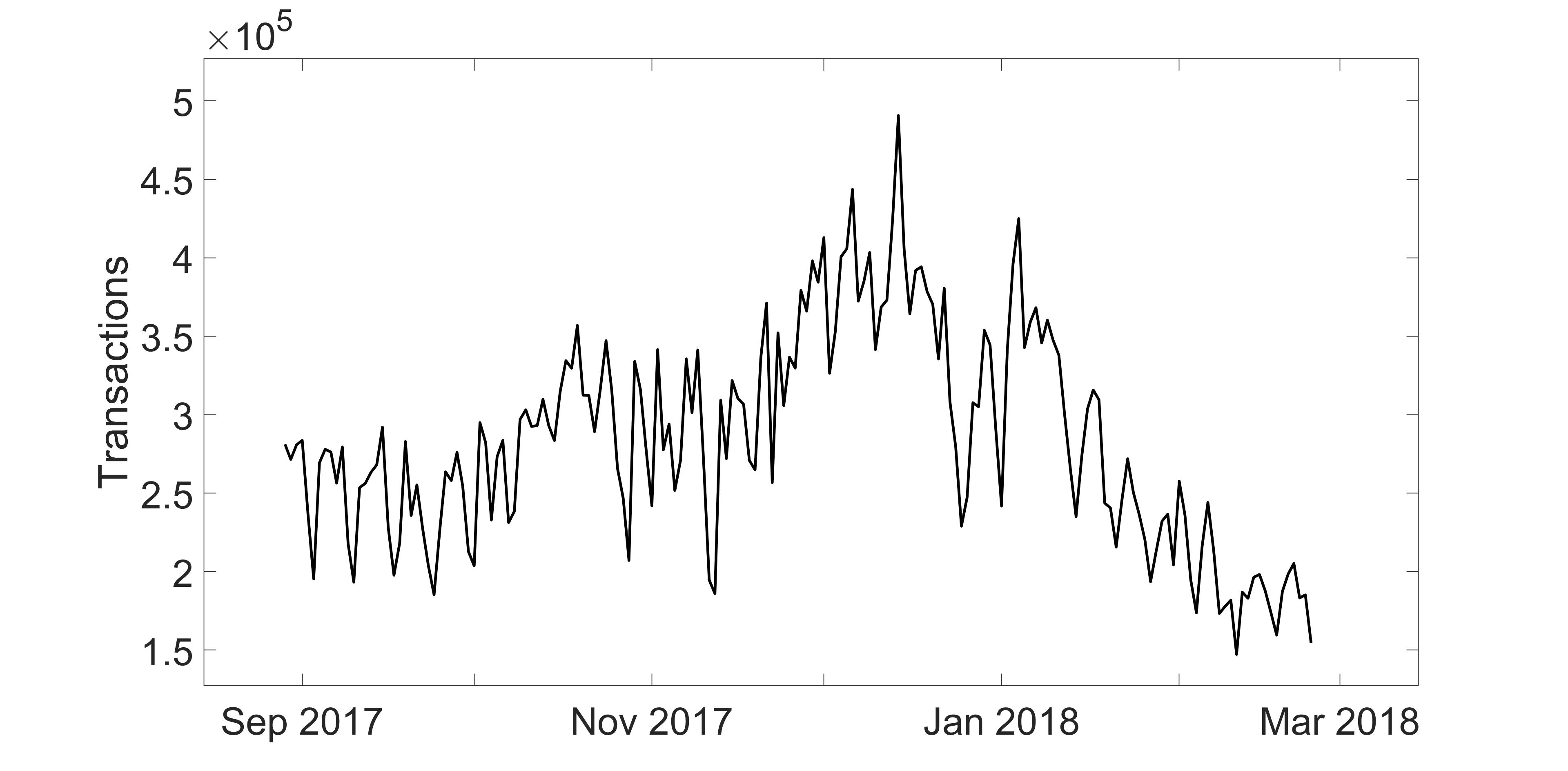}}
    \caption{Confirmed transactions per day (based on data from \cite{txsperday})}
    \label{fig:txsperday}
\end{figure}

In the next two experiments, we consider a node fluctuating on and off the network after joining with an empty mempool. Alongside the node with intermittent connectivity, we have a second Linux machine that also starts with an empty mempool, but with subsequently stable network connectivity.  The two experiments took place during the Spring of $2018$. As seen in Figures~\ref{fig:mempooltxs} and~\ref{fig:txsperday}, the mempool of the stable node usually had $5{,}000$ to $10{,}000$ transactions and the network was experiencing around $150{,}000$ to $200{,}000$ transactions per day incoming, as compared to the mempools having between $10{,}000$ and $100{,}000$ transactions with $300{,}000$ to $350{,}000$ transactions per day incoming in Fall $2017$.

\subsubsection{Experimental Setup}
To simulate a node fluctuating on and off the network, we install Bitcoin Core v0.15.0 on a Linux machine (Ubuntu 16.04 LTS) and run a Python script that turns the networking on and off with a $90\%$ uptime with $10$ minute intervals. Each interval consists of the node being online for $9$ minutes and offline for $1$ minute. ``Empty'' mempools are established for both nodes in the same manner as described in Section~\ref{subsec:intermittent} and the logging system is used to record the data as well as how FalafelSync's performance. For the third experiment (presented in the next section), we also record what transactions are sent in the FalafelSync messages and  the state of the mempool at the time. Our scripts are available at~\cite{pythonrepo}.

\subsubsection{Results}

\begin{figure}[!t]
	\centering
    \makebox[\textwidth][c]{\includegraphics[width=1.1\textwidth]{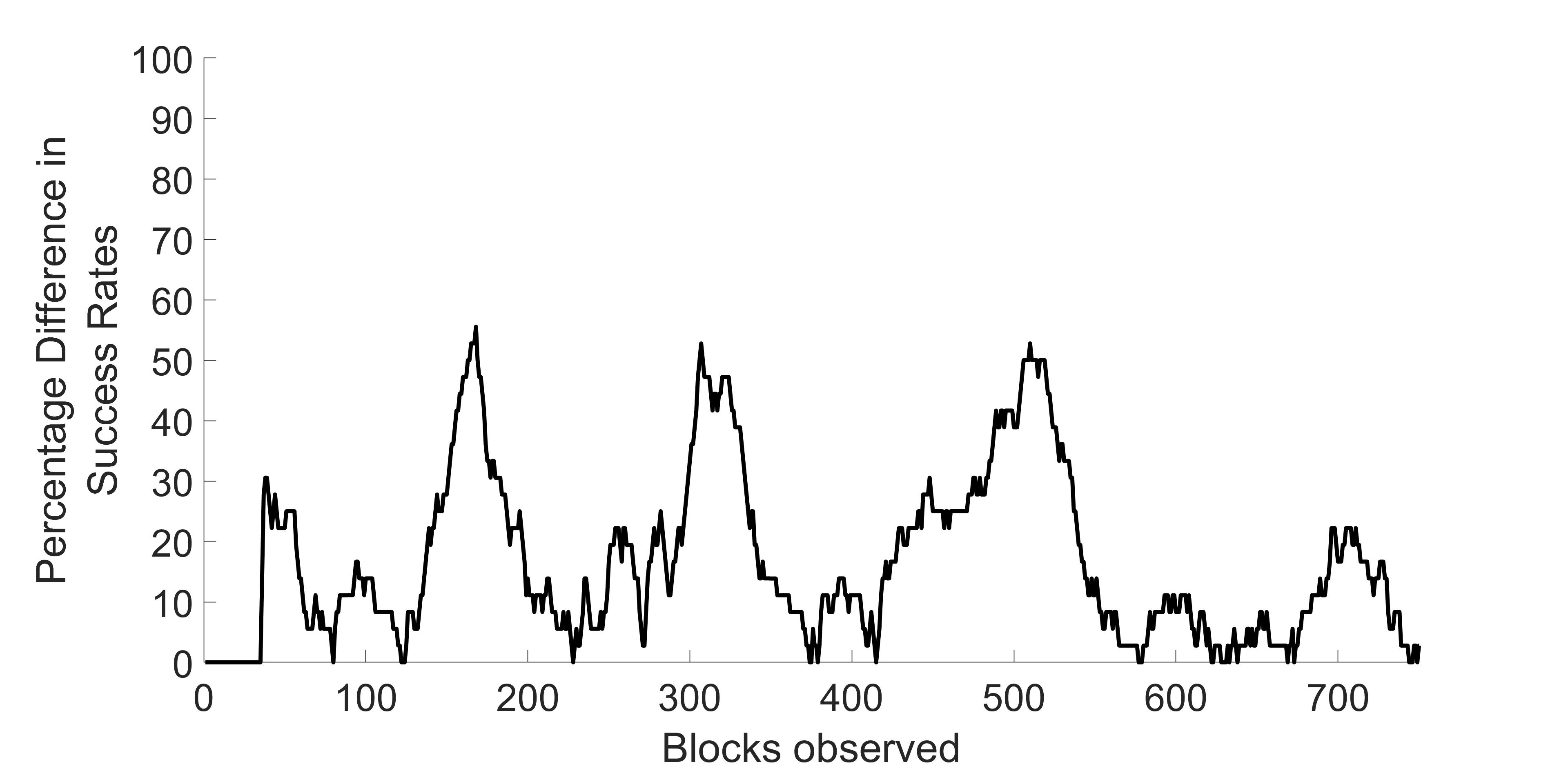}}
    \caption{Compact block success rate without FalafelSync}
    \label{fig:compsuccessrate_withoutsync}
\end{figure}

\begin{figure}[t]
	\centering
    \makebox[\textwidth][c]{\includegraphics[width=1.1\textwidth]{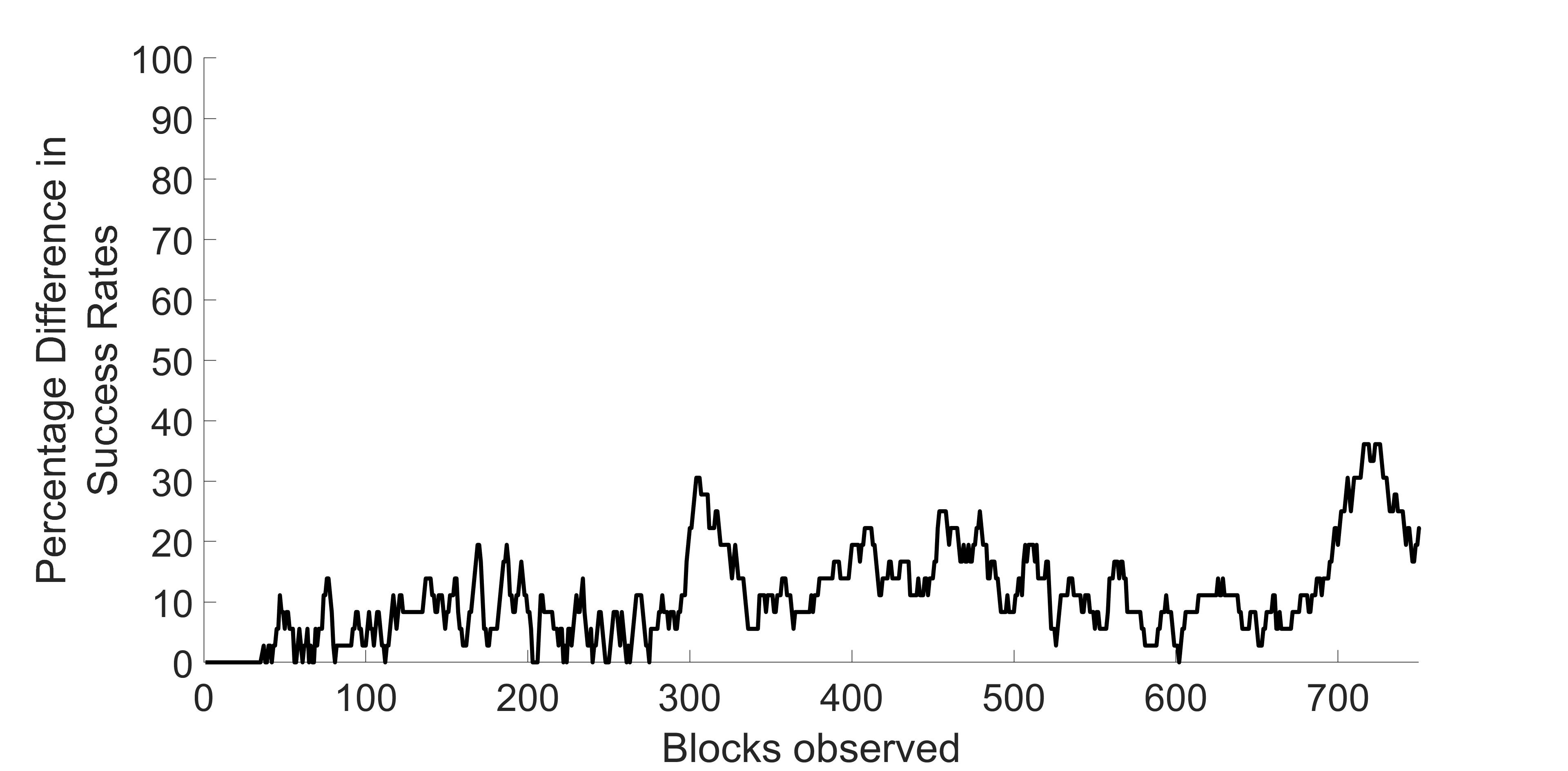}}
    \caption{Compact block success rate with FalafelSync}
    \label{fig:compsuccessrate_withsync}
\end{figure}

In Figure~\ref{fig:compsuccessrate_withoutsync}, we observe the percentage point difference between the performance of the stable node and intermittent node for a given window. For the intermittent node, we see that a node sometimes performed as well as the stable node but was not able to recover from the intermittent networking. This result is not a particularly surprising one, but it is important to take note of the significant number of compact blocks failing and the inability of this node to recover from the fluctuations. 

From the data we collected, in the second experiment, the stable node recovers in around $120$ blocks. This result may seem like a significant improvement over the previous experiment, but that is not the case when taking into consideration the difference in activity between the networks of the two experiments. In our first experiment, mempools contained a magnitude more transactions than in the second and third experiments; this means that the block-time it takes to overturn the mempool is about a magnitude longer. With this in mind we can see that both nodes recovered at the same rate when adjusting for network activity. 

\subsection{Intermittent Network Connectivity with FalafelSync}
%Other ideas: "Testing FalafelSync" , "Effect of FalafelSync on nodes with empty mempool and intermittent network connection"

\begin{figure}[!b]
	\centering
    \makebox[\textwidth][c]{\includegraphics[width=1\textwidth]{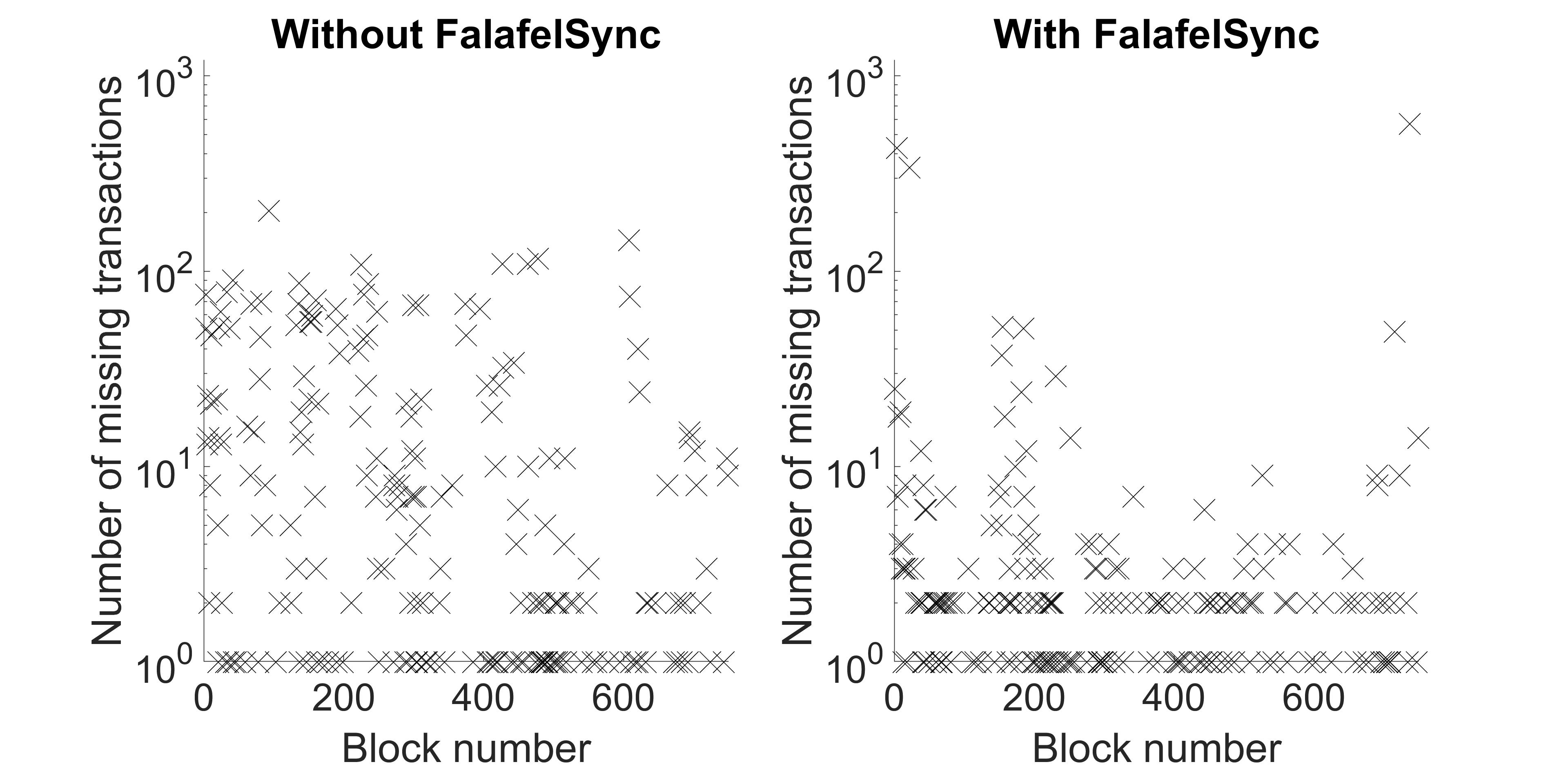}}
    \caption{Comparison of the number of missing transactions without and with FalafelSync on an intermittently connected node}
    \label{fig:comp_node003}
\end{figure}

For our third round of experiments, we set up both nodes in a similar fashion to the second experiment. However, the nodes partially sync their mempools with each other at regular intervals every ten minutes, with some natural deviations due to the use of the pseudo-timer, using the FalafelSync described earlier. Comparing Figures~\ref{fig:compsuccessrate_withoutsync} and~\ref{fig:compsuccessrate_withsync} we can see an immediate improvement for the node with intermittent network connection. It performs much more closely to the stable node in terms of compact block success rate and missing transactions for failed compact blocks. It takes the stable node around $300$ blocks to recover and the intermittent node about $700$ blocks, although after block $700$ we do see a small divergence in performance between the stable node and intermittent node. This could be the case that the missing transactions are received when the intermittent node is offline. Figure~\ref{fig:comp_node003} shows that a node running without FalafelSync misses several hundred transactions. With FalafelSync implemented, the count of missing transactions reduces significantly.

\begin{table}[t]
\centering
\caption{Results collected from the first 750 blocks observed}
\label{my-label}
\begin{tabular}{c|c|c|c|c|}
\cline{2-5}
                                            & \multicolumn{2}{c|}{No Sync} & \multicolumn{2}{c|}{Sync}   \\ \cline{2-5} 
                                            & Stable       & Intermittent      & Stable      & Intermittent      \\ \hline
\multicolumn{1}{|c|}{Success} & & & & \\
\multicolumn{1}{|c|}{Compact} & 661           & 544          & 627          & 542          \\ 
\multicolumn{1}{|c|}{Block} & & & & \\\hline
\multicolumn{1}{|c|}{Average} & & & & \\
\multicolumn{1}{|c|}{Success}     & 89.32\%        & 72.53\%       & 84.73\%       & 72.27\%       \\ 
\multicolumn{1}{|c|}{Rate} & & & & \\\hline
\multicolumn{1}{|c|}{Difference} & \multicolumn{2}{c|}{}  & \multicolumn{2}{c|}{} \\
\multicolumn{1}{|c|}{In Success}   & \multicolumn{2}{c|}{16.79\%}  & \multicolumn{2}{c|}{12.46\%} \\ 
\multicolumn{1}{|c|}{Rate} & \multicolumn{2}{c|}{}  & \multicolumn{2}{c|}{}\\\hline
\end{tabular}
\end{table}

With our implementation of FalafelSync, we are able to narrow down the performance difference between the nodes with stable and intermittent network connection by $4.4\%$ from a raw performance index perspective. While this result may seem small, it is a $25\%$ improvement over the performance differences between the former and latter node's performance without FalafelSync.

\subsection{Simulations of Graphene with FalafelSync}

With the FalafelSync in place and observed improvement in the success rate of compact blocks, we analyze the data collected by the logging system from previously mentioned experiment runs to find out how the new synchronization protocol affects the success rate of the Graphene block. For our simulations we considered a Graphene block to fail if more than $15\%$ transactions in the Graphene block are not in the receiving node's mempool as that is a large enough difference for the block's IBLT to not decode successfully. 

\begin{figure}[t]
	\centering
    \includegraphics[width=0.85\textwidth]{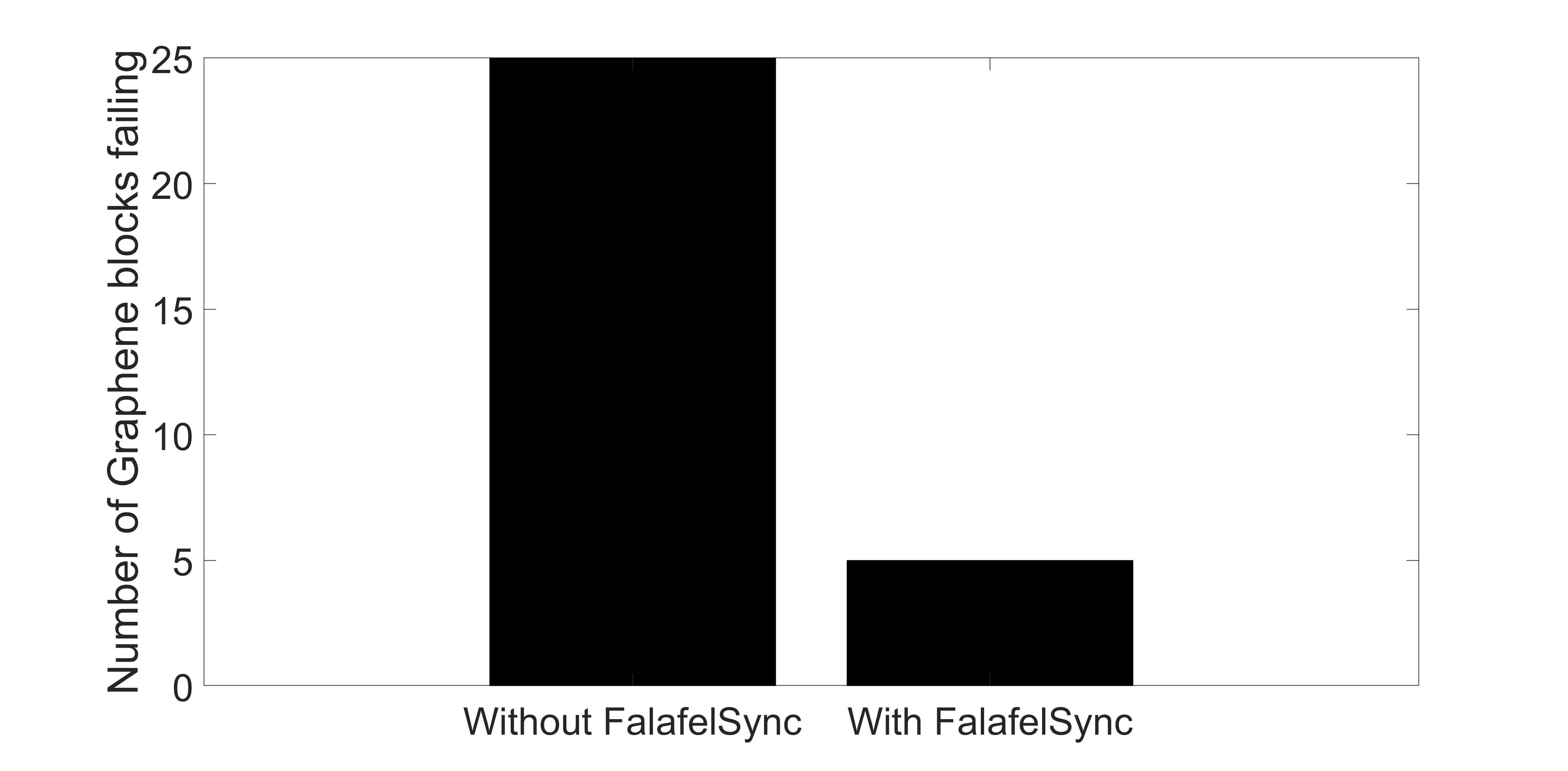}
    \caption{Number of Graphene block IBLTs that fail to decode on an intermittently connected node}
    \label{fig:graphenefailingblocks}
\end{figure}

\begin{figure}[t]
	\centering
    \includegraphics[width=0.85\textwidth]{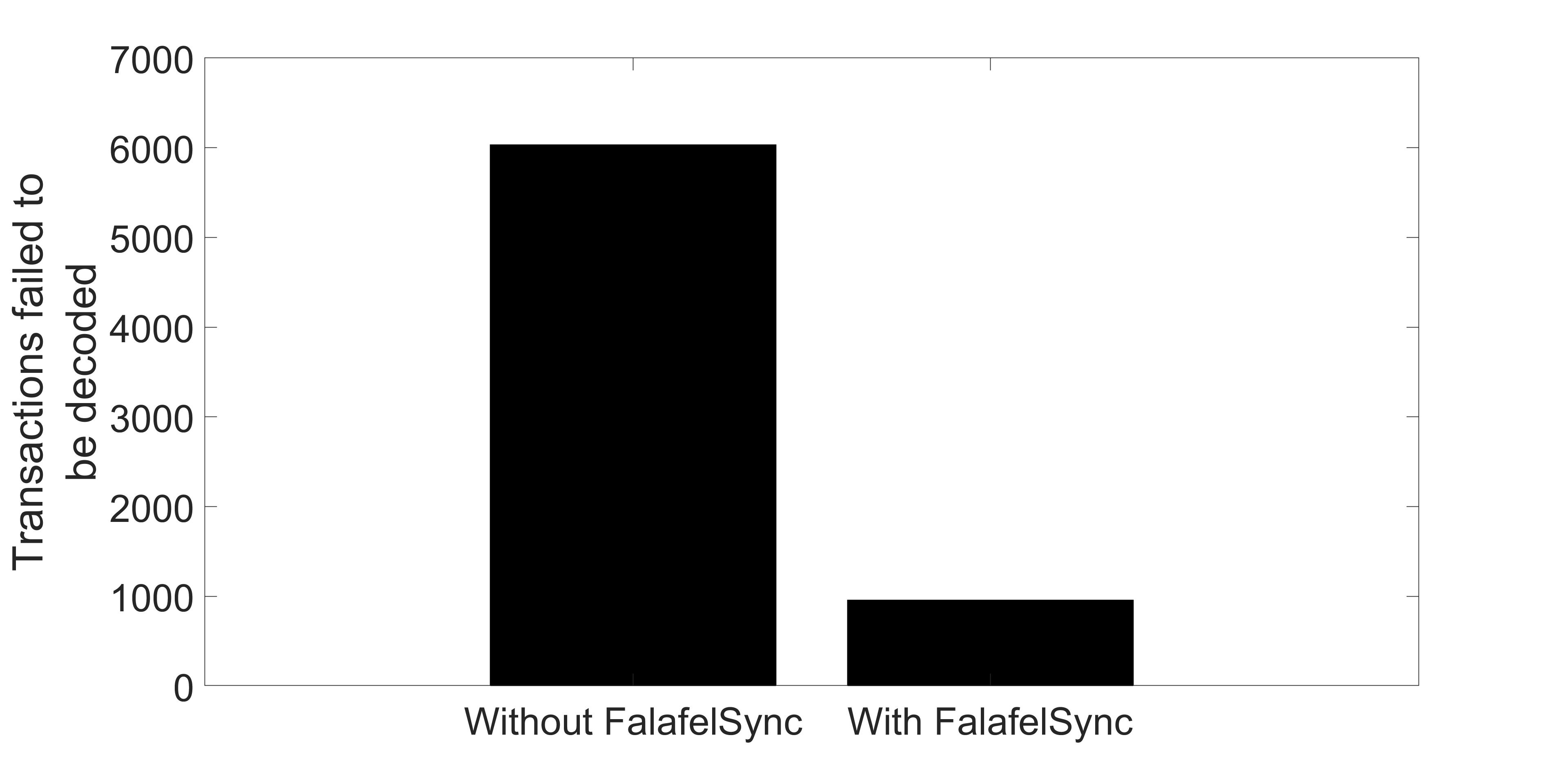}
    \caption{Number of transactions in Graphene blocks that failed to decode without and with FalafelSync}
    \label{fig:grapheneTXsmissing}
\end{figure}

Figures~\ref{fig:graphenefailingblocks} and \ref{fig:grapheneTXsmissing} illustrate our simulation results. The first Figure shows that a node without FalafelSync implemented experienced $5$ times more failed Graphene blocks as compared to the node with FalafelSync implemented. The second Figure shows that the node implementing FalafelSync is able to successfully decode transactions with Graphene at a rate 6 times higher than a node that does not implement FalafelSync. This is because the intermittent node had fewer missing transactions in its mempool because of synchronization and the IBLT could be decoded.

%looks good 


\subsection{Overhead caused by FalafelSync}

It is important to note that FalafelSync does impose some overhead on the standard Bitcoin protocol. As discussed earlier, the number of transactions sent by FalafelSync in an \texttt{inv} message is the larger of the top $10\%$ of transactions in the mempool and $1000$ transactions in the mempool, sorted according to \texttt{ancestor-score}. For each transaction in the \texttt{inv} message, $32$ bytes are transmitted. The node receiving the sync message then responds with a \texttt{getdata} message, which is a list of indexes of the \texttt{inv} message that are not present in its mempool. This ends up taking roughly $4$ bytes per transaction requested. The last message sent is a full copy of each requested transaction, and this can be between $500$ to $800$ bytes per transaction. We note that FalafelSync does not save bandwidth when compared to the Compact block's fall-back procedure; however it does save round-trip time (and associated latency) by preparing the node for the next coming compact block.

\subsection{Limitations}

When working with Bitcoin, several factors can affect the results in unforeseen ways. In particular, Bitcoin's topology is a somewhat dense directed random graph, meaning that not all information may reach a node if there is no efficient path from the source. This can cause a stable node to miss transactions due to the topology of the network. 

All of our experiments are conducted with two nodes (running on dedicated computers) that initially have empty mempools. It would be desirable to test the sync protocol with additional nodes that are never taken offline. To further test the effectiveness of FalafelSync, running several experiments in parallel to make sure they have the same network activity could help eliminate some of the randomnesses in the measured performance. Running  experiments of FalafelSync over longer periods (say, beyond two  weeks) would also be useful to gain higher statistical confidence in the results.

%==============CONCLUSION=============
\section{Conclusion}
\label{sec:conclusions}

In this paper, we  document a previously unknown phenomenon in Bitcoin's network, namely its limited resilience to churn. Our experiments show that the performance of the compact block protocol and, by extension, the Graphene protocol significantly degrades when a node is joining the network or when a node has intermittent connectivity. We present and implement FalafelSync, a proof-of-concept synchronization protocol, to improve the resilience of the Bitcoin network in such cases. Concurrently, we develop a logging system that facilitates performing measurements on the Bitcoin network and characterizing its behavior.

%================FUTURE WORK============
%\section{Future Works}

Moving forward, collecting statistics from additional nodes, located in different geographic areas would be valuable to confirm the general applicability of the results presented in this paper. Running the experiments under different levels of network activity, and especially when the network is overloaded, would help further assess the positive impact of FalafelSync in minimizing the recovery times of nodes joining the network. 

A practical solution to prioritized transaction synchronization will most likely involve a combination of sophisticated algorithms for finding differences and new data-structures for reconciliation.  Potential candidates for prioritized transaction synchronization include IBLT~\cite{goodrich2011invertible} for their high tolerance for differences and CPISync~\cite{trachtenberg2002fast,starobinski2003efficient,jin2012prioritized} for protocols with near-optimal communication complexity. Attempting to implement a full-fledged system of synchronization algorithms and data structures will be challenging but worthwhile for the longterm development of Bitcoin and, more generally, any distributed application that relies on a blockchain. 

\section*{Acknowledgment}
This work was supported in part by the NSF under grant CCF-1563753. Any opinions, findings, and conclusions or recommendations expressed in this material are those of the authors and do not necessarily reflect the views of the NSF.

%\medskip
\bibliographystyle{unsrt}
\bibliography{bibliography}

\begin{thebibliography}{10}

\bibitem{article}
Satoshi Nakamoto.
\newblock Bitcoin: A peer-to-peer electronic cash system.
\newblock 03 2009.

\bibitem{bitnodes}
Bitnodes.
\newblock \url{https://bitnodes.earn.com/dashboard/}, 2018.
\newblock Online; Accessed: February 24, 2018.

\bibitem{cryptocurrencies}
Cryptocurrency market capitalizations.
\newblock \url{https://coinmarketcap.com/all/views/all/}, 2018.
\newblock Online; Accessed: February 24, 2018.

\bibitem{decker2013information}
Christian Decker and Roger Wattenhofer.
\newblock Information propagation in the bitcoin network.
\newblock In {\em Peer-to-Peer Computing (P2P), 2013 IEEE Thirteenth
  International Conference on}, pages 1--10. IEEE, 2013.

\bibitem{bitcoinwikiblock}
Block.
\newblock \url{https://en.bitcoin.it/wiki/Block}, 2016.
\newblock Online; Accessed: February 24, 2018.

\bibitem{energy}
Bitcoin energy consumption index.
\newblock \url{https://digiconomist.net/bitcoin-energy-consumption}, 2018.
\newblock Online; Accessed: February 24, 2018.

\bibitem{BIP152}
Matt Corallo.
\newblock Compact block relay.
\newblock \url{https://github.com/bitcoin/bips/blob/master/bip-0152.mediawiki},
  2016.

\bibitem{ozisik2017graphene}
A~Pinar Ozisik, Gavin Andresen, George Bissias, Amir Houmansadr, and Brian
  Levine.
\newblock Graphene: A new protocol for block propagation using set
  reconciliation.
\newblock In {\em Data Privacy Management, Cryptocurrencies and Blockchain
  Technology}, pages 420--428. Springer, 2017.

\bibitem{stutzbach2006understanding}
Daniel Stutzbach and Reza Rejaie.
\newblock Understanding churn in peer-to-peer networks.
\newblock In {\em Proceedings of the 6th ACM SIGCOMM conference on Internet
  measurement}, pages 189--202. ACM, 2006.

\bibitem{ourbitcoinrepo}
Nabeel Younis.
\newblock Bitcoin.
\newblock \url{https://github.com/Nabeelperson/bitcoin}, 2017.

\bibitem{avgtxs}
Average number of transactions per block.
\newblock
  \url{https://blockchain.info/charts/n-transactions-per-block?timespan=2years},
  2017.
\newblock Online; Accessed: February 25, 2018.

\bibitem{txspersec}
Transaction rate.
\newblock \url{https://blockchain.info/charts/transactions-per-second}, 2017.
\newblock Online; Accessed: February 26, 2018.

\bibitem{blocksize}
Block size limit controversy.
\newblock \url{https://en.bitcoin.it/wiki/Block\_size\_limit\_controversy},
  2017.
\newblock Online; Accessed: February 24, 2018.

\bibitem{protodocs}
Protocol documentation.
\newblock \url{https://en.bitcoin.it/wiki/Protocol\_documentation}, 2018.
\newblock Online; Accessed: February 24, 2018.

\bibitem{goodrich2011invertible}
Michael~T Goodrich and Michael Mitzenmacher.
\newblock Invertible bloom lookup tables.
\newblock In {\em Communication, Control, and Computing (Allerton), 2011 49th
  Annual Allerton Conference on}, pages 792--799. IEEE, 2011.

\bibitem{bitcoincore}
Bitcoin.
\newblock \url{https://github.com/bitcoin/bitcoin/tree/v0.15.0}, 2017.
\newblock Online; Accessed: February 24, 2018.

\bibitem{bitcoindevref}
Bitcoin developer reference.
\newblock
  \url{https://bitcoin.org/en/developer-reference\#remote-procedure-calls-rpcs},
  2017.
\newblock Online; Accessed: February 24, 2018.

\bibitem{minsky2003set}
Yaron Minsky, Ari Trachtenberg, and Richard Zippel.
\newblock Set reconciliation with nearly optimal communication complexity.
\newblock {\em IEEE Transactions on Information Theory}, 49(9):2213--2218,
  2003.

\bibitem{mempooltxs}
Mempool transaction count.
\newblock \url{https://blockchain.info/charts/mempool-count?timespan=1year},
  2017.
\newblock Online; Accessed: February 25, 2018.

\bibitem{txsperday}
Confirmed transactions per day.
\newblock \url{https://blockchain.info/charts/n-transactions?timespan=180days},
  2017.
\newblock Online; Accessed: February 25, 2018.

\bibitem{pythonrepo}
Nabeel Younis.
\newblock Falafelsyncpythonscripts.
\newblock \url{https://github.com/Nabeelperson/FalafelSyncPythonScripts}, 2018.

\bibitem{trachtenberg2002fast}
Ari Trachtenberg, David Starobinski, and Sachin Agarwal.
\newblock Fast pda synchronization using characteristic polynomial
  interpolation.
\newblock In {\em INFOCOM 2002. Twenty-First Annual Joint Conference of the
  IEEE Computer and Communications Societies. Proceedings. IEEE}, volume~3,
  pages 1510--1519. IEEE, 2002.

\bibitem{starobinski2003efficient}
David Starobinski, Ari Trachtenberg, and Sachin Agarwal.
\newblock Efficient pda synchronization.
\newblock {\em IEEE Transactions on Mobile Computing}, 2(1):40--51, 2003.

\bibitem{jin2012prioritized}
Jiaxi Jin, Wei Si, David Starobinski, and Ari Trachtenberg.
\newblock Prioritized data synchronization for disruption tolerant networks.
\newblock In {\em MILITARY COMMUNICATIONS CONFERENCE, 2012-MILCOM 2012}, pages
  1--8. IEEE, 2012.

\end{thebibliography}
\end{document}